\documentclass[twocolumn]{aastex631}

\usepackage{amsmath}	
\usepackage{threeparttable} 
\usepackage{xspace}
\usepackage{rotating}
\usepackage{booktabs}

\newcommand{\fluxunit}{$10^{-17}$ erg s$^{-1}$ cm$^{-2}$ \AA$^{-1}$}

\newcommand{\hetg}{$g_\mathrm{HETDEX}$}
\newcommand{\OII}{[\ion{O}{2}]\xspace}
\newcommand{\lya}{Ly$\alpha$\xspace}

\received{March 4, 2025}
\revised{}
\accepted{}

\shorttitle{DESI Spectroscopy of HETDEX Detections}
\shortauthors{Landriau et al.}

\begin{document}

\title{DESI Spectroscopy of HETDEX Emission-line Candidates I: Line Discrimination Validation}

\correspondingauthor{Martin Landriau}
\email{mlandriau@lbl.gov}

\author[0000-0003-1838-8528]{Martin Landriau}
\affiliation{Lawrence Berkeley National Laboratory,
1 Cyclotron Road, Berkeley, CA 94720, USA}

\author[0000-0002-2307-0146]{Erin Mentuch Cooper}
\affiliation{Department of Astronomy, The University of Texas at Austin, Austin, TX 78712, USA}
\affiliation{McDonald Observatory, The University of Texas at Austin, 2515 Speedway Boulevard, Austin, TX 78712, USA}

\author[0000-0002-8925-9769]{Dustin Davis}
\affiliation{Department of Astronomy, The University of Texas at Austin, Austin, TX 78712, USA}

\author[0000-0002-8433-8185]{Karl Gebhardt}
\affiliation{Department of Astronomy, The University of Texas at Austin, Austin, TX 78712, USA}

\author[0000-0002-1328-0211]{Robin Ciardullo} \affiliation{Department of Astronomy and Astrophysics, The Pennsylvania
State University, University Park, PA 16802, USA}
\affiliation{Institute for Gravitation and the Cosmos, The Pennsylvania State University, University Park, PA 16802, USA}

\author[0000-0001-7600-5148]{\'Eric Armengaud}
\affiliation{Saclay - Commissariat à l'\'energie atomique et aux \'energies alternatives,
IRFU, CEA, Universit\'{e} Paris-Saclay, F-91191 Gif-sur-Yvette, France}

\author[0000-0002-4928-4003]{Arjun Dey}
\affiliation{NSF NOIRLab, 950 N. Cherry Ave., Tucson, AZ 85719, USA}

\author[0000-0001-5999-7923]{Anand Raichoor}
\affiliation{Lawrence Berkeley National Laboratory,
1 Cyclotron Road, Berkeley, CA 94720, USA}

\author{David J. Schlegel}
\affiliation{Lawrence Berkeley National Laboratory,
1 Cyclotron Road, Berkeley, CA 94720, USA}

\author{Michael Wilson}
\affiliation{Department of Astrophysical Sciences, Princeton University, Peyton Hall, Princeton, NJ 08544, USA}


\author{J.~Aguilar}
\affiliation{Lawrence Berkeley National Laboratory,
1 Cyclotron Road, Berkeley, CA 94720, USA}

\author[0000-0001-6098-7247]{S.~Ahlen}
\affiliation{Physics Dept., Boston University, 590 Commonwealth Avenue, Boston, MA 02215, USA}

\author[0000-0001-9712-0006]{D.~Bianchi}
\affiliation{Dipartimento di Fisica ``Aldo Pontremoli'', Universit\`a degli Studi di Milano, Via Celoria 16, I-20133 Milano, Italy"}
\affiliation{INAF-Osservatorio Astronomico di Brera, Via Brera 28, 20122 Milano, Italy}

\author{D.~Brooks}
\affiliation{Department of Physics \& Astronomy, University College London, Gower Street, London, WC1E 6BT, UK}

\author{T.~Claybaugh}
\affiliation{Lawrence Berkeley National Laboratory, 1 Cyclotron Road, Berkeley, CA 94720, USA}

\author[0000-0002-1769-1640]{A.~de la Macorra}
\affiliation{Instituto de F\'{\i}sica, Universidad Nacional Aut\'{o}noma de M\'{e}xico,  Circuito de la Investigaci\'{o}n Cient\'{\i}fica, Ciudad Universitaria, Cd. de M\'{e}xico  C.~P.~04510,  M\'{e}xico}

\author[0000-0003-4992-7854]{S.~Ferraro}
\affiliation{Lawrence Berkeley National Laboratory, 1 Cyclotron Road, Berkeley, CA 94720, USA}
\affiliation{University of California, Berkeley, 110 Sproul Hall \#5800 Berkeley, CA 94720, USA}

\author[0000-0002-2890-3725]{J.~E.~Forero-Romero}
\affiliation{Departamento de F\'isica, Universidad de los Andes, Cra. 1 No. 18A-10, Edificio Ip, CP 111711, Bogot\'a, Colombia}
\affiliation{Observatorio Astron\'omico, Universidad de los Andes, Cra. 1 No. 18A-10, Edificio H, CP 111711 Bogot\'a, Colombia}

\author{E.~Gazta\~{n}aga}
\affiliation{Institut d'Estudis Espacials de Catalunya (IEEC), c/ Esteve Terradas 1, Edifici RDIT, Campus PMT-UPC, 08860 Castelldefels, Spain}
\affiliation{Institute of Cosmology and Gravitation, University of Portsmouth, Dennis Sciama Building, Portsmouth, PO1 3FX, UK}
\affiliation{Institute of Space Sciences, ICE-CSIC, Campus UAB, Carrer de Can Magrans s/n, 08913 Bellaterra, Barcelona, Spain}

\author[0000-0003-3142-233X]{S.~Gontcho A Gontcho}
\affiliation{Lawrence Berkeley National Laboratory, 1 Cyclotron Road, Berkeley, CA 94720, USA}
\affiliation{University of Virginia, Department of Astronomy, Charlottesville, VA 22904, USA}

\author{G.~Gutierrez}
\affiliation{Fermi National Accelerator Laboratory, PO Box 500, Batavia, IL 60510, USA}

\author[0000-0003-1197-0902]{C.~Hahn}
\affiliation{Steward Observatory, University of Arizona, 933 N, Cherry Ave, Tucson, AZ 85721, USA}

\author[0000-0002-6550-2023]{K.~Honscheid}
\affiliation{Center for Cosmology and AstroParticle Physics, The Ohio State University, 191 West Woodruff Avenue, Columbus, OH 43210, USA}
\affiliation{Department of Physics, The Ohio State University, 191 West Woodruff Avenue, Columbus, OH 43210, USA}

\author[0000-0002-1081-9410]{C.~Howlett}
\affiliation{School of Mathematics and Physics, University of Queensland, Brisbane, QLD 4072, Australia}

\author[0000-0002-6024-466X]{M.~Ishak}
\affiliation{Department of Physics, The University of Texas at Dallas, 800 W. Campbell Rd., Richardson, TX 75080, USA}

\author[0000-0002-0000-2394]{S.~Juneau}
\affiliation{NSF NOIRLab, 950 N. Cherry Ave., Tucson, AZ 85719, USA}

\author{R.~Kehoe}
\affiliation{Department of Physics, Southern Methodist University, 3215 Daniel Avenue, Dallas, TX 75275, USA}

\author[0000-0003-3510-7134]{T.~Kisner}
\affiliation{Lawrence Berkeley National Laboratory, 1 Cyclotron Road, Berkeley, CA 94720, USA}

\author[0000-0001-6356-7424]{Anthony Kremin}
\affiliation{Lawrence Berkeley National Laboratory, 1 Cyclotron Road, Berkeley, CA 94720, USA}

\author[0000-0001-7178-8868]{L.~Le~Guillou}
\affiliation{Sorbonne Universit\'{e}, CNRS/IN2P3, Laboratoire de Physique Nucl\'{e}aire et de Hautes Energies (LPNHE), FR-75005 Paris, France}

\author[0000-0003-1887-1018]{Michael E.~Levi}
\affiliation{Lawrence Berkeley National Laboratory, 1 Cyclotron Road, Berkeley, CA 94720, USA}

\author[0000-0003-4962-8934]{M.~Manera}
\affiliation{Departament de F\'{i}sica, Serra H\'{u}nter, Universitat Aut\`{o}noma de Barcelona, 08193 Bellaterra (Barcelona), Spain}
\affiliation{Institut de F\'{i}sica d'Altes Energies (IFAE), The Barcelona Institute of Science and Technology, Edifici Cn, Campus UAB, 08193, Bellaterra (Barcelona), Spain}

\author[0000-0002-1125-7384]{A.~Meisner}
\affiliation{NSF NOIRLab, 950 N. Cherry Ave., Tucson, AZ 85719, USA}

\author{R.~Miquel}
\affiliation{Instituci\'{o} Catalana de Recerca i Estudis Avan\c{c}ats, Passeig de Llu\'{\i}s Companys, 23, 08010 Barcelona, Spain}
\affiliation{Institut de F\'{i}sica d'Altes Energies (IFAE), The Barcelona Institute of Science and Technology, Edifici Cn, Campus UAB, 08193, Bellaterra (Barcelona), Spain}

\author[0000-0002-2733-4559]{J.~Moustakas}
\affiliation{Department of Physics and Astronomy, Siena College, 515 Loudon Road, Loudonville, NY 12211, USA}

\author[0000-0001-9070-3102]{S.~Nadathur}
\affiliation{Institute of Cosmology and Gravitation, University of Portsmouth, Dennis Sciama Building, Portsmouth, PO1 3FX, UK}

\author[0000-0001-6979-0125]{I.~P\'erez-R\`afols}
\affiliation{Departament de F\'isica, EEBE, Universitat Polit\`ecnica de Catalunya, c/Eduard Maristany 10, 08930 Barcelona, Spain}

\author{C.~Poppett}
\affiliation{Lawrence Berkeley National Laboratory, 1 Cyclotron Road, Berkeley, CA 94720, USA}
\affiliation{Space Sciences Laboratory, University of California, Berkeley, 7 Gauss Way, Berkeley, CA  94720, USA}

\author[0000-0001-7145-8674]{F.~Prada}
\affiliation{Instituto de Astrof\'{i}sica de Andaluc\'{i}a (CSIC), Glorieta de la Astronom\'{i}a, s/n, E-18008 Granada, Spain}

\author{G.~Rossi}
\affiliation{Department of Physics and Astronomy, Sejong University, 209 Neungdong-ro, Gwangjin-gu, Seoul 05006, Republic of Korea}

\author[0000-0002-9646-8198]{E.~Sanchez}
\affiliation{CIEMAT, Avenida Complutense 40, E-28040 Madrid, Spain}

\author{M.~Schubnell}
\affiliation{Department of Physics, University of Michigan, 450 Church Street, Ann Arbor, MI 48109, USA}

\author{D.~Sprayberry}
\affiliation{NSF NOIRLab, 950 N. Cherry Ave., Tucson, AZ 85719, USA}

\author[0000-0003-1704-0781]{G.~Tarl\'{e}}
\affiliation{University of Michigan, 500 S. State Street, Ann Arbor, MI 48109, USA}

\author{B.~A.~Weaver}
\affiliation{NSF NOIRLab, 950 N. Cherry Ave., Tucson, AZ 85719, USA}

\author[0000-0001-5381-4372]{R.~Zhou}
\affiliation{Lawrence Berkeley National Laboratory, 1 Cyclotron Road, Berkeley, CA 94720, USA}

\author[0000-0002-6684-3997]{H.~Zou}
\affiliation{National Astronomical Observatories, Chinese Academy of Sciences, A20 Datun Rd., Chaoyang District, Beijing, 100012, P.R. China}

\author[0000-0003-2575-0652]{Daniel J. Farrow}
\affiliation{Centre of Excellence for Data Science, Artificial Intelligence \& Modelling (DAIM),\\ University of Hull, Cottingham Road, Hull, HU6 7RX, UK}
\affiliation{E. A. Milne Centre for Astrophysics, University of Hull, Cottingham Road, Hull, HU6 7RX, UK}

\author[0000-0001-6717-7685]{Gary J. Hill}
\affiliation{Department of Astronomy, The University of Texas at Austin, Austin, TX 78712, USA}
\affiliation{McDonald Observatory, The University of Texas at Austin, 2515 Speedway Boulevard, Austin, TX 78712, USA}

\author[ 0000-0002-8434-979X]{Donghui Jeong}
\affiliation{Department of Astronomy and Astrophysics, The Pennsylvania State University, University Park, PA 16802, USA}
\affiliation{Institute for Gravitation and the Cosmos, The Pennsylvania State University, University Park, PA 16802, USA}

\author[0000-0001-5561-2010]{Chenxu Liu}
\affiliation{South-Western Institute for Astronomy Research, Yunnan University, Kunming, Yunnan, 650500, People’s Republic of China}

\author[0000-0002-6186-5476]{Shun Saito}
\affiliation{Institute for Multi-messenger Astrophysics and Cosmology, Department of Physics, Missouri University of Science and Technology, 1315 N Pine St, Rolla, MO 65409}
\affiliation{Kavli Institute for the Physics and Mathematics of the Universe (WPI), Todai Institutes for Advanced Study, the University of Tokyo, Kashiwanoha, Kashiwa, Chiba 277-8583, Japan}

\author[0000-0001-7240-7449]{Donald P. Schneider}
\affiliation{Department of Astronomy and Astrophysics, The Pennsylvania State University, University Park, PA 16802, USA}
\affiliation{Institute for Gravitation and the Cosmos, The Pennsylvania State University, University Park, PA 16802, USA}



\begin{abstract}

The Hobby-Eberly Dark Energy Experiment (HETDEX) is an untargeted spectroscopic galaxy survey that uses \lya emitting galaxies (LAEs) as tracers of $1.9 < z < 3.5$ large scale structure.  Most detections consist of a single emission line, whose identity is inferred via a Bayesian analysis of ancillary data.  To determine the accuracy of these line identifications, HETDEX detections were observed with the Dark Energy Spectroscopic Instrument (DESI\null).  In two DESI pointings, high confidence spectroscopic redshifts are obtained for 1157 sources, including 982 LAEs. The DESI spectra are used to evaluate the accuracy of the HETDEX object classifications, and tune the methodology to achieve the HETDEX science requirement of $\lesssim 2\%$ contamination of the LAE sample by low-redshift emission-line galaxies, while still assigning 96\% of the true Ly$\alpha$ emission sample with the correct spectroscopic redshift.  We compare emission line measurements between the two experiments assuming a simple Gaussian line fitting model. Fitted values for the central wavelength of the emission line, the measured line flux and line widths are consistent between the surveys within uncertainties. Derived spectroscopic redshifts, from the two classification pipelines, when both agree as an LAE classification, are consistent to within $\langle \Delta z / (1 + z) \rangle = 6.9\times 10^{-5}$ with an rms scatter of $3.3\times 10^{-4}$.

\vspace{5pt}
Data available at \texttt{\url{https://data.desi.lbl.gov/desi/public/dr1/vac/dr1/hetdex}}.

\end{abstract}



\section{Introduction}
\label{sec:intro}

HETDEX \citep[]{HETDEXinst, HETDEXsurvey} is an untargeted galaxy survey that
uses \lya emitting galaxies (LAEs) as tracers of large-scale structure in the redshift
range $1.88 < z < 3.52$ .  The survey
instrument, the Visible Integral-Field Replicable Unit Spectrograph (VIRUS) consists of 78 integral field units (IFUs), each with 448 $1\farcs 5$- diameter fibers, feeding  156 spectrographs with a spectral resolution of 5.6~\AA ~ ($R\sim$800) in the wavelength range 3500 - 5500~\AA\null.  The goal of HETDEX is to measure the power spectrum of LAEs and thereby constrain dark energy at $z \sim 2$ with as little contamination as possible from foreground objects. 

The challenge of HETDEX is that normal LAEs, i.e. those without AGN emission, are single emission-line sources in HETDEX spectra.  However, because of the limited wavelength range of the VIRUS spectrographs, foreground [\ion{O}{2}] emitting galaxies between $0.13 < z < 0.47$ also appear as single-line objects.  Since the resolution of VIRUS is insufficient to resolve the [\ion{O}{2}] doublet, these objects can be mistaken for LAEs.  Misclassifications are therefore a serious issue, since to be successful, HETDEX needs to ensure that contamination of the LAE sample by [\ion{O}{2}] galaxies is $\lesssim 2\%$ \citep{HETDEXsurvey}.

During a pilot survey, \citet{PilotSurvey1} demonstrated that a equivalent width (EW) cuts can be used to separate LAEs from most [\ion{O}{2}] galaxies.  (See also \citealp{Gawiser2007}.)  By supplementing the EW selection using a modified Bayesian approach based on prior knowledge of the galaxy populations, \citep{leung2017, davis2023} show that that the misclassification of [\ion{O}{2}] galaxies in the HETDEX LAE sample may be as low as 1.5\%.  This method uses the measured distribution functions of line luminosity and EWs of LAEs and local Universe [OII] emitters in order to compute the posterior probability that a single emission line object is an LAE or [OII] emitter (see references in \citep{leung2017} for a list of data used).  However, their conclusion was based on a comparison sample of spectroscopically confirmed objects that was biased towards brighter high-redshift galaxies.

\begin{figure*}[th]
\includegraphics[width=\textwidth]{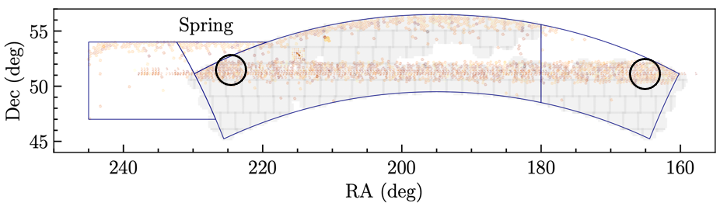}
  \caption{HETDEX main (a.k.a ``spring'') field outlined in blue, with completion status as indicated by orange tiles as of January 2021 when the target list for the sample was selected (the grey tiles were planned but yet to be observed).  The two circles show the position of the two
    DESI tiles chosen for the follow-up.}\label{fig:fields}
\end{figure*}

The Dark Energy Spectroscopic Instrument (DESI)~\citep{Snowmass2013.Levi, DESI2016a.Science, DESI2016b.Instr,DESI2022.KP1.Instr, FocalPlane.Silber.2023, Corrector.Miller.2023, FiberSystem.Poppett.2024}, survey is a program of targeted spectroscopy aimed at building the largest 3D map of the Universe to date, and thereby inferring the expansion history of the Universe up to redshift $z=3.5$.  The instrument consists of 5000
robot-positioned fibers feeding 10 3-arm spectrographs covering the
wavelength range 3600-9800\,\AA . With its wider wavelength coverage and higher spectral resolution ($R$ = 2,000 - 3,200) relative to HETDEX, DESI spectra have a clear advantage over HETDEX for the classification of faint emission-line objects.  Not only can DESI resolve the [\ion{O}{2}] doublet, but its longer exposure times and wider wavelength range can detect continuum emission and reveal additional spectral lines such as H$\alpha$ for lower-redshift sources, and \ion{C}{3]} $\lambda 1909$ and \ion{C}{4} in higher-redshift objects.

The focus of this paper is to
determine whether the methodology of \citet{leung2017} as implemented by \citet{davis2023} does indeed meet the HETDEX science requirements for low-redshift contamination.  A second goal of the analysis is to be part of a pilot survey for DESI-2, a proposed survey targeting high-$z$ galaxies using DESI; this aspect will be
studied in a companion paper.

Section~\ref{sec:ts} describes our target selection process and the DESI observations.  Section~\ref{sec:analysis} presents our analysis and the results of our visual classification and redshift determinations.  In Section~\ref{sec:classification} the DESI spectroscopic sample is used to measure HETDEX's contamination rate and accuracy. Finally, Section~\ref{sec:linecomps} presents a comparison of measured line fluxes between the two experiments; and Section~\ref{sec:conclusion} states our conclusions. All wavelengths stated in this paper are as measured in air unless otherwise noted. No dust corrections are applied, although a measure of visual dust extinction, $A_{V}$ is included in the catalog provided with this paper to be applied as desired. A full description of the data accompanying the release is provided in the Appendix.

\section{Target selection and observations}\label{sec:ts}

During the DESI survey validation phase (SV)~\citep{DESI2023a.KP1.SV}, a number of secondary
target projects were carried out, one of them being a follow up of
HETDEX detections.
Figure~\ref{fig:fields} show the HETDEX main survey footprint and the completion
status at the time of designing the target selection.  The circles show the two DESI pointings.   At the time of target selection, HETDEX was in its early stages and these fields had the most complete coverage within the area of the DESI focal plane and contained a number of
interesting objects.

The parent sample of HETDEX detections are from HDR2 (internal data release version 2.1.2). A subset of this dataset is presented in the HETDEX Public Source Catalog 1 \citep{cooper2023}, the difference being that the public release was limited to emission lines detections with a signal-to-noise ratio ($S/N$) above 5.5 (the internal sample is limited to sources with $S/N\ge4.8$) and contain some omissions due to quality assessment. In this paper, we cross-match the input sample to those in a later HETDEX internal release (HDR3) to better calibrate with the latest HETDEX reduction pipeline. Note: separations between the input coordinates (listed by `TARGET\_RA', `TARGET\_DEC') and the HETDEX positions are provided in the column `SEP' in the data table released with this paper. The initial target coordinates are consistent with the later catalog release (DR3) within 0\farcs3.

The DESI sample was sub-selected from the main HETDEX emission line galaxy catalog based on two criteria:

\begin{enumerate}
    \item \textit{LAE candidates}: A \hetg\ value (a pseudo $g$-magnitude measured from the HETDEX spectrum itself) fainter than 22.5~mag (AB) and a P(\lya) likelihood value greater than 0.5 \citep{leung2017}. Line emitters with P(\lya) greater than 0.5 are assigned a redshift assuming the detected emission line is \lya.  Conversely, detections with P(\lya) less than 0.5 are generally assigned a redshift assuming the emission line is due to [\ion{O}{2}] $\lambda 3727$.  (Other observables, such as the detection of a second emission line, are also incorporated into the probability value; for more details, see \citealt{davis2023}.)

    \item \textit{Borderline LAE/[\ion{O}{2}] emitters:} Faint galaxies with 
    \hetg\ value between 23 and 25~mag as well as a P(\lya) likelihood value between 0.3 and 0.5 in the second date release HETDEX catalog.
    This sub-sample represents galaxies  which are either faint [\ion{O}{2}] emitters or bright (LBG-like) LAEs and will be most difficult to classify based on a single emission line. This selection does bias the sample towards slightly brighter targets. As a result the contamination that is measured from low redshift emission line interlopers is expected to be biased slightly higher compared to the full HETDEX sample.
  
\end{enumerate}

\begin{figure}[t]
  \includegraphics[width=0.45\textwidth]{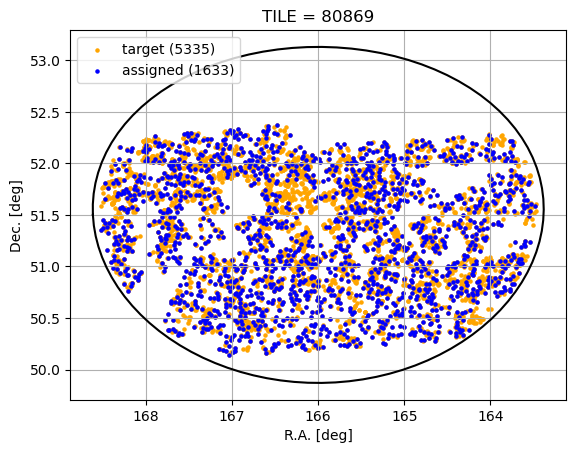}

  \includegraphics[width=0.45\textwidth]{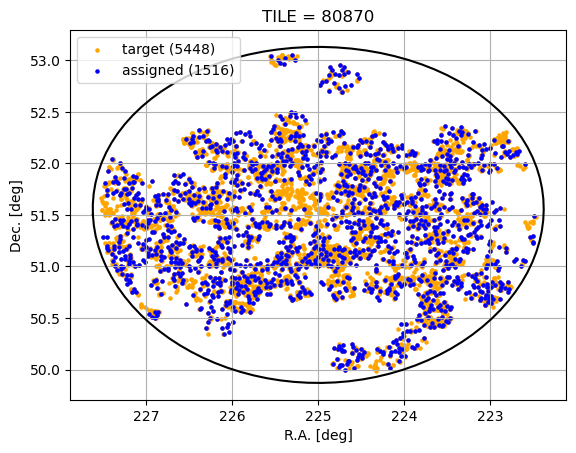}
  \caption{Parent and assigned target distributions for the two pointings. The numbers 80869 and 80870 refer to the DESI tileID, which is unique for the combination of pointing and fiber assignments. Gaps in the assigned target distribution where there were available targets in the parent sample are due to fibers being assigned to other secondary target programs or small regions of the focal plane with non-working fiber positioning robots during that period.}\label{fig:fiberassign}
\end{figure}

Based on these criteria, HETDEX provided DESI with 10,780 possible targets.  DESI then positioned fibers on 1633 (out of 5335) in Pointing 1 and 1516 (out of 5448) objects in Pointing 2, for a total of 3149 targets. This selection was based on fiber positioning requirements and fiber availability.  The distribution of selected targets with valid fibers for the two pointings is shown in Figure~\ref{fig:fiberassign}.

The data were taken between 13 March and 13 May 2021 and are
summarized in Table~\ref{tab:exp}.  Note that the total effective exposure times of 18944.0~sec for tile 80869 and 11306.4~sec for tile 80870 are considerably longer than the nominal 1000~seconds used for regular DESI dark time tiles (though some targets
have less accumulated time, due to fiber placement inaccuracies in some exposures). The reason for the cumulative exposure for tile 80869 was $\sim 60\%$ longer than that for tile 80870 is that time was initially allocated for only one of the two fields; towards the end of the survey validation period
\citep[SV;][]{DESI2023a.KP1.SV,DESI2023b.KP1.EDR}, 
it was decided to expand the secondary target programs as much as time allowed before starting the main survey.  The spectra obtained through these observations are available as part of the DESI first data release (DR1)~\citep{DESI_DR1,DESI2024.II.KP3,2024arXiv240403000D,2024arXiv240403001D,DESI2024.V.KP5,2024arXiv240403002D,DESI2024.VII.KP7B}.

\begin{table*}\label{tab:exp}
  \begin{center}
\begin{tabular}{l|l||l|l|l|l|l}
 DESI tileID & RA (deg) & DEC (deg) & $N_\textrm{targets}$ & N$_\textrm{nights}$ & Exposures (sec) &ExpTime$_\textrm{eff}$ (sec) \\
  \hline
  \hline
&  &  &  & & $2\times 60$ & \\
80869 & 165 & 51.5 & 1959 & 4 & $9\times 900$ & 18944.0\\
            & & & & & $7\times 1220$ & \\
  \hline
  &  &  &  & & $1\times 900$ & \\
80870 & 225 & 51.5 & 1826 & 3 & $2\times 1800 $ & 11306.4\\
&  &  &  & & $2\times 1820$ & \\
\end{tabular}\caption{Summary of exposures. The exposures column lists the number and duration of the exposures on the respective fields, and ExpTime$_{\textrm{eff}}$ denotes the total effective exposure time.}
\end{center}
\end{table*}

\section{Visual Inspection}\label{sec:analysis}

The raw data were processed using the DESI spectroscopic pipeline~\citep{SurveyOps.Schlafly.2023,Spectro.Pipeline.Guy.2023} and every spectra were visually inspected using the DESI VI tool
\texttt{prospect}\footnote{\texttt{\url{https://github.com/desihub/prospect/}}}.
 Sources were
classified into five categories, numbered 0 to 4, based on the confidence for which the
redshift could be determined as follows:
\begin{enumerate}
    \item[4] Confident classification: two or more secure features.
    \item[3] Probable classification: at least one secure spectral feature and continuum or many weak spectral features.
    \item[2] Possible classification: one strong spectral feature but unsure what it is.
    \item[1] Unlikely classification: clear signal but features are unidentified.
    \item[0] Nothing there, no signal.
\end{enumerate}
For more details of the DESI visual inspection procedure and results from SV, see \citet{VIGalaxies.Lan.2023} and \citet{VIQSO.Alexander.2023}.

The analysis presented here used the data reductions of Data Release 1. 

\begin{figure*}[ht]
\centering
\includegraphics[width=0.9\textwidth]{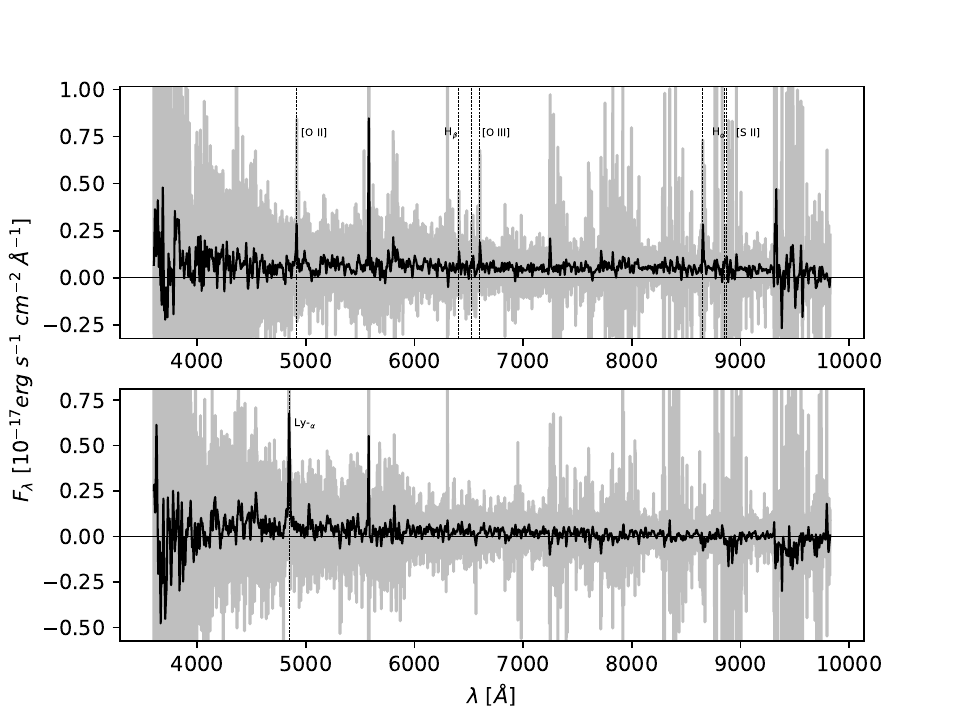}
\caption{Example DESI spectra of a faint [\ion{O}{2}] galaxy at $z=0.318$ (top) and an LAE at $z=2.984$ (bottom) confirmed by visual
  inspection.}
  \label{fig:LAEexamples}
\end{figure*}

\subsection{Preliminary cuts}

After excluding spectra due to fiber-positioning errors, several additional cuts were made before determining the accuracy of the HETDEX
line classification algorithm.

Some HETDEX
sources had their assumed \lya line too close to the blue end of the DESI spectral
range for an effective confirmation.  We therefore excluded those HETDEX objects whose emission line wavelength was $\lambda \le 3610$~\AA\null.  A further restriction was imposed to require HETDEX sources to have good IFU coverage as there is evidence for an increase in the HETDEX false positive rate for objects located near the edges of an IFU (\texttt{APCOR\_DEX} $> 0.6$, where \texttt{APCOR\_DEX} is the fraction of area covered by VIRUS fibers within a 3\farcs5 aperture radius centered on the detection).  We also restricted the sample to sources that were in both the final HETDEX second data release and in the third release, which used new, improved reduction algorithms.   After these criteria were applied, our final sample contained 2374 sources, of which 1157 (49.0\%) were detected with confidence by DESI, i.e., with \texttt{VI\_QUALITY} $\ge 3$.

\subsection{Visual Inspection}

The DESI automated redshift assignment algorithm proved to be quite poor for most HETDEX emission-line objects. Since the majority of the sample does not have significant continuum emission, many of the spectral template fits either failed or found spurious emission lines in the red.  The result was generally an erroneous redshift, even when it was returned with a high confidence. Thus, redshifts are assigned based on visual inspection. If multiple line identifications are visible, a highly confident redshift can be assigned to the source.  This is the case for low-redshift objects, where multiple spectral lines are available in the DESI spectral range. An example of such a source can be seen in the top panel of Figure~\ref{fig:LAEexamples}. In addition, for low-redshift, emission-line sources, a zoom in on the [\ion{O}{2}] doublet reveals a double peak in the DESI spectra providing a high confidence redshift (\texttt{VI\_QUALITY}=4).

In contrast, the majority of HETDEX LAE candidates are single emission line sources with no continua. An example is given in the bottom panel of Figure~\ref{fig:LAEexamples}. Even with the increased sensitivity of the DESI spectrograph, the continuum level of the LAEs is not detected. No absorption features are available to secure the redshift and, in most cases, no other emission lines are available. If no other (lower redshift) solution fits the spectrum,  we assume the emission line is due to \lya, and we tune the redshift accordingly. In the case of a single emission line source, we assign a \texttt{VI\_QUALITY} of 3. If other emission lines are detected, such as \ion{C}{4} or \ion{C}{3}], then a \texttt{VI\_QUALITY} of 4 is assigned. In the case of a very noisy spectrum in which a peak in emission matches the HETDEX redshift, we tune the \lya line to this peak and assign a lower \texttt{VI\_QUALITY} of 1 or 2 depending on the relative strength of the line to the noise. Emission line galaxies blueward of 3800~\AA\ are difficult to confirm with DESI spectroscopy due to a number of spiky noise features.

Two independent classifications were made on each member of the sample. When a disagreement in classification occurred, a re-assessment was made on the source such that both classifiers were in agreement. Out of over 2500 classifications, only 15 re-assessments were made.

\subsection{Results}

\begin{table}[t]

 \begin{center} \begin{tabular}{l|r|r|r}
    Type  &VI quality$\ge$0   & VI quality$\ge 3$ & VI quality$\ge$4\\
    \hline
    \hline
  ALL  &   2374     &     1157    &         303 \\
 LAE  &   2077     &      982    &         137 \\
 \null [\ion{O}{2}]   &    241     &      131    &         131 \\
 AGN  &     36     &       32    &          24 \\
 Other &     17    &       12    &          11 \\
    \end{tabular}
  \end{center}
  \caption{Summary of results of visual inspection}\label{tab:sumres}
  \end{table}
  
The results of our HETDEX-DESI comparison are summarized in Table~\ref{tab:sumres}. Objects are separated into different groups, labeled \texttt{SOURCE\_TYPE} in the catalog accompanying this paper, based on the spectroscopic redshifts obtained from Visual Inspection (\texttt{VI\_Z}).  For objects without spectroscopic redshifts from VI (\texttt{VI\_Z}=-1, \texttt{VI\_QUALITY}=0), the HETDEX catalog indicates the source type as described in \citet{cooper2023}. The object groups are as follows: 

\begin{itemize}
\item LAE: A $\texttt{VI\_Z} > 1.9$ emission-line source not previously identified as an AGN in the HETDEX AGN catalog \citep{liu2022}.
\item AGN: An emission-line source previously identified as an AGN candidate in \citet{liu2022}.
\item OII: An [\ion{O}{2}] emission-line galaxy at $z < 0.5$.  
\item Other: A confirmed redshift of a non-emission line source or an intermediate redshift object with emission lines beyond the HETDEX spectral range of 3500--5500~\AA\null. These are usually blended sources that do not confirm the HETDEX emission-line detection, because it is not possible to tell if a line belongs to the foreground and background sources, and thus infer proper classification (see \citep{cooper2023} for details).
\end{itemize}

All [\ion{O}{2}] sources are confirmed with high confidence (\texttt{VI\_QUALITY}=4). In all spectra, other emission lines (such as H$\alpha$, H$\beta$, and [\ion{O}{3}]) confirm the spectroscopic redshift and the [\ion{O}{2}] doublet at 3727.092\,\AA\ and 3729.875\,\AA\ is resolved. For the AGN sample, 24/32 sources are confirmed with multiple emission lines. Those that are confirmed with only one broad line and do not show other spectral features and are assigned \texttt{VI\_QUALITY}=3. These are candidate Lyman Alpha Blobs  (LABs) that were misidentified as AGN due to the broad nature of the line. They exhibit extended \lya emission and little to no continua (more information on LABs in HETDEX is presented in Mentuch Cooper et al., in prep).

A number of LAE candidates (137/982) are confirmed with high confidence based on the presence of additional spectral lines (generally \ion{C}{4} and \ion{C}{3}]). Although not previously identified as AGN, these sources exhibit signatures of an AGN contribution. Either the additional lines are relatively faint or they fall outside the spectral window of the VIRUS spectrographs.

\begin{figure}
\includegraphics[width=0.5\textwidth]{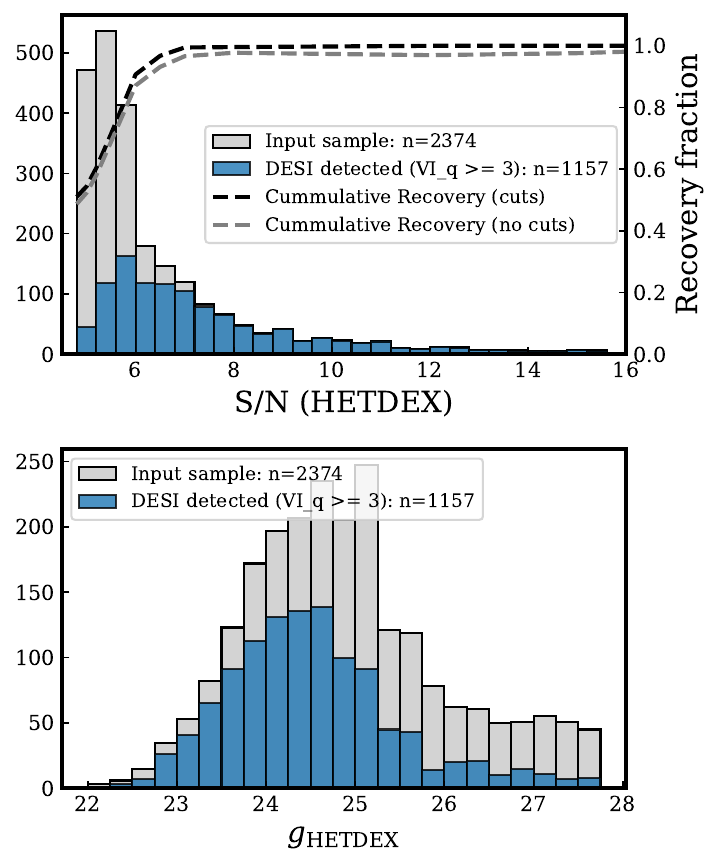}
\caption{The top panel shows the distribution of emission-line detection $S/N$ for HETDEX targets. The grey histogram indicates the range in the input target sample, while the blue shows the recovered galaxies with VI$\_$quality$\ge3$. The cumulative recovery fraction as a function of signal-to-noise ratio is shown in dashed grey for the full sample and in dashed black line where some cuts are applied. Only well positioned sources with emission lines $\lambda>3800$ are included in this sample and those that have not been excluded from the sample in later HETDEX catalogs (ie. we only include those emission lines with \texttt{DEX\_FLAG==1}). Sources at low $S/N$ are missed due to astrometric uncertainties for low $S/N$ sources and possible false positives. The bottom panels presents the distribution in \hetg\ magnitudes of the input and recovered sample, in grey and blue respectively.
}\label{fig:sn_dist}
\end{figure}

Figure~\ref{fig:sn_dist} shows the distribution of sources from the input target sample (grey) and the recovered sample (blue) as a function of $S/N$ and \hetg\ magnitude on the top and bottom panels, respectively. In the top panel, we include the cumulative recovery fraction of sources (confirmed with \texttt{VI\_QUALITY}$\ge3$ above a given $S/N$ threshold). In grey, we plot the cummulative recovery without any sub-selection of the sample, while in dashed-black, we select only those targets with emission lines redward of 3800~\AA\ which are not flagged by HETDEX for various reasons (i.e., we exclude internally flagged detections, \texttt{DEX\_FLAG}$=0$, objects).

The distribution of visual mangitude is provided in the bottom panel of Figure~\ref{fig:sn_dist}. The targeted sample of LAEs is biased towards slightly higher \hetg\ values compared to the LAE sample presented in the HETDEX public catalog in \citet{cooper2023}. A sub-sample of brighter LAE candidates were selected in this study to test the HETDEX classification method on LBG-like \lya emitting galaxies that can potentially be confused with fainter, low-redshift [\ion{O}{2}]-emitters.

\subsection{Missing sources}
\label{sec:missing}

At high HETDEX signal-to-noise ($S/N$ $> 6.5$), 97\% (448/463) of the sample is recovered when we account for only those sources in the latest HETDEX catalog (\texttt{DEX\_FLAG==1}), whose TARGET fiber position is well matched to the HDR3 detection (\texttt{SEP}$<0\farcs3$) and
whose emission line is at a wavelength greater
than 3800\,\AA\null.
After visual inspection, roughly half the sources are likely false positives in the HETDEX sample. Most of these are higher line width objects, \textit{i.e.} emission lines with a line width greater than 6\AA\, that are not identified as a representative source or are not included in the AGN catalog that are excluded in most HETDEX analyses in order to remove detections that are identified by the HETDEX detection pipeline due to sharp discontinuities in the detection spectrum that do not correspond to true line emission.

However, four of these objects are missed by DESI simply because they happen to be related to a brighter resolved object, but not centered on that object. The HETDEX detection search algorithm described in \citet{HETDEXsurvey} is designed to search for point sources, but in the case of extended objects, multiple detections can arise for the same source. Bright point sources such as stars and AGN also result in multiple HETDEX detections. When the target list was created, detection grouping to merge a set of multiple redundant detections into single HETDEX source \citep[as described in][]{cooper2023}, was not yet developed. For this reason, some DESI fibers were not centered on their HETDEX target. With its single fiber and better image quality, DESI missed the object.

At lower HETDEX $S/N$ the number of missing sources increases. Specifically, at a $S/N$ threshold of 6, 93\% of HETDEX sources are recovered by DESI, but by $S/N$=5.5,
the cumulative recovery rate drops to 75\% and to 56\% at S/N=5.0.
Simulations show that due to the large ($1\farcs 5$-diameter) size of the HETDEX fibers and fiber-to-fiber throughput variations, the computed centroid of a low $S/N$ source may be offset from its true position on the sky by over 1\arcsec.  Specifically, the simulations described in \citet{HETDEXsurvey}, as well as comparison to independent emission line samples, show that at $S/N>5.5$, the median centering accuracy of a HETDEX source is about 0\farcs4 with up to 10\% of sources recovered at over 1\arcsec\ away. Moreover, at lower S/N, this astrometric error quickly increases, so that at the limit of HETDEX detections, the median centering accuracy is 0\farcs6, and over 20\% of sources are located more than 1\arcsec\ away from their input position. As the DESI spectrograph uses single $1\farcs 5$ fibers and has better image quality than HETDEX, it is likely that if positioned off center from the true position of an LAE, the signal would be lost completely.

False positive contamination in the HETDEX sample also contributes to the increase in missing sources at lower HETDEX $S/N$\null.  At $S/N$ values above 5.5, false detections in the HETDEX database are less than 10\% \citep{cooper2023}.  However, when integrated over all S/N, the false positive rate is likely between 15--30\%\citep{HETDEXsurvey}. Unfortunately, due to the centering uncertainties described above, the DESI spectroscopy cannot be used to better define this value. More generally, DESI cannot say anything about these missing sources and are not used in the subsequent analysis.

\section{HETDEX Classification}
\label{sec:classification}

The details of the spectral classification methodologies employed by HETDEX are presented in \cite{davis2023} and \cite{cooper2023}, and only a brief overview is provide here. 

For the HETDEX wavelength window and depth, often only a single emission line is visible: generally \lya ($1.9 < z < 3.5$) or \OII ($0.13 < z < 0.5$). Moreover, because the resolution of the VIRUS spectrograph is only $\lambda$/$\Delta \lambda \sim 800$, the \OII doublet is not resolvable in the HETDEX data \citep[][]{HETDEXsurvey,davis2023}. It is therefore necessary to use on statistical methods to distinguish these lines.

The primary classification code that performs this analysis is the Emission Line eXplorer \citep[ELiXer]{davis2023} which builds on the Bayesian analysis in \cite{leung2017} with  modifications in \cite{Farrow+2021}.  ELiXer is a significant improvement over the commonly used \lya-restframe 20~\AA\ equivalent width threshold for \lya and \OII emission lines \citep[][]{Gronwall2007a,Adams2011a}, as it takes advantage of additional information, both in the HETDEX spectra and, when available, in external photometric and spectroscopic catalogs.  In brief, ELiXer enhances the simple 20~\AA\ $W_{Ly\alpha}$ cut with a weighted voting scheme. This scheme incorporates various ``votes'' for or against individual or sets of possible classifications based on different elements. Each vote is assigned a weight derived from the frequency with which it aligns with the correct classification, given a training set of spectra with known redshifts. The votes are summed with their weights to determine the likely classification in reference to a configurable P(\lya) threshold value, a normalized value between 0 and 1 (where 0 = not likely \lya~ and 1 = likely \lya). Note that P(\lya), while expressed in probability terms and has a value between $0$ and $1$, is not a probability in the formal sense. In addition, a set of specialized conditions are evaluated and may modify the redshift or enhance the classification \citep[Section 3 of][]{davis2023}.

As one purpose of this work is to evaluate the performance of the HETDEX spectral classifications on their own merit, the use of all external phot-$z$ and spec-$z$ catalogs is disabled. Imaging cutouts from external photometric archives are still used for bandpass aperture fluxes, as are object sizes as measured by the ELiXer code. These are the same testing conditions imposed in \cite{davis2023}.

\begin{figure}[t]
    \centering
        \hspace*{-0.5cm}
\includegraphics[width=0.5\textwidth]{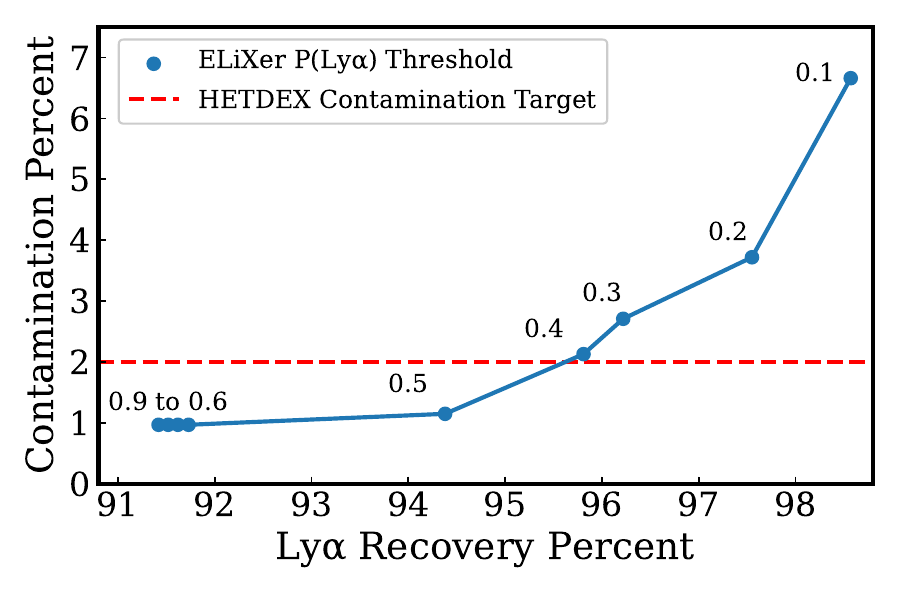}
\caption{The results of our comparison of ELiXer classifications of HETDEX spectra with the high confidence visually vetted DESI spectra. The $y$-axis gives the percent contamination of the HETDEX LAE sample by misidentified \OII galaxies, while the percentage of LAEs recovered by the classifier is on the $x$-axis. The labeled points show the corresponding P(\lya) threshold used to configure the ELiXer classifier. The red horizontal dashed line represents the requirements of the HETDEX survey \citep{HETDEXsurvey}, though higher contamination rates may be acceptable with the use of additional decontamination methodologies \citep{Farrow+2021}. Despite differences in the redshift and magnitude distributions with the Spectrocopic-$z$ Assessment Sample (SzAS) in \cite{davis2023}, the curve is very similar.
} \label{fig:elixer_plya}
\end{figure}

The contamination and recovery fractions of the LAE sample as a function of the P(\lya) threshold are presented in Table~\ref{tab:elixer_plya} and Figure~\ref{fig:elixer_plya}. The data clearly show the trade-off between the two parameters. Here, we define contamination as the number of \OII emission lines that were incorrectly classified by ELiXer as \lya divided by the total number of \lya lines (as classified by ELiXer). \OII $\lambda 3727$ is the most likely contaminant in the HETDEX \lya sample and therefore has the most impact on the measurement of the $z \sim 2.5$ galaxy clustering signal \citep{HETDEXsurvey,davis2023}. The recovery fraction (represented as a percent in the figure) is the number of correctly classified \lya lines, again divided by the total number of \lya lines in the sample.  As described above, this test sample is limited to the visually vetted subset of objects with \texttt{VI\_QUALITY} $\ge 3$.

The results shown in Figure \ref{fig:elixer_plya} are consistent with those in \cite{davis2023}, and supports the efficacy and robustness of the HETDEX classification mechanism. In particular, the distribution of the objects in the earlier work: (\textit{1}) skews towards much brighter objects, in which all sources are in the range $g=$ 22.0--25.5, with more than 85\% of sources being brighter than $g=23.5$ (see Figure 7 in \cite{davis2023} and Figure \ref{fig:sn_dist} in this work) and (\textit{2}) has a smaller fraction of \lya sources (24\% in \cite{davis2023} vs.\ 85\% here).

The default P(\lya) setting of 0.5 yields a good recovery of \lya sources with a contamination fraction due to low-redshift sources mis-classified as \lya below the 2\% science requirement specified by HETDEX \citep{HETDEXsurvey}. With some additional decontamination methodologies, HETDEX may be able to sustain larger contamination, perhaps even above 5\% \citep{Farrow+2021}.  If so, there is opportunity to move to a lower P(\lya) threshold and recover additional \lya sources at the expense of a modest increase in contamination. 

\begin{figure}[t]
  \includegraphics[trim={0.1cm 0.4cm 0.5cm 1.5cm},clip, width=0.5\textwidth]{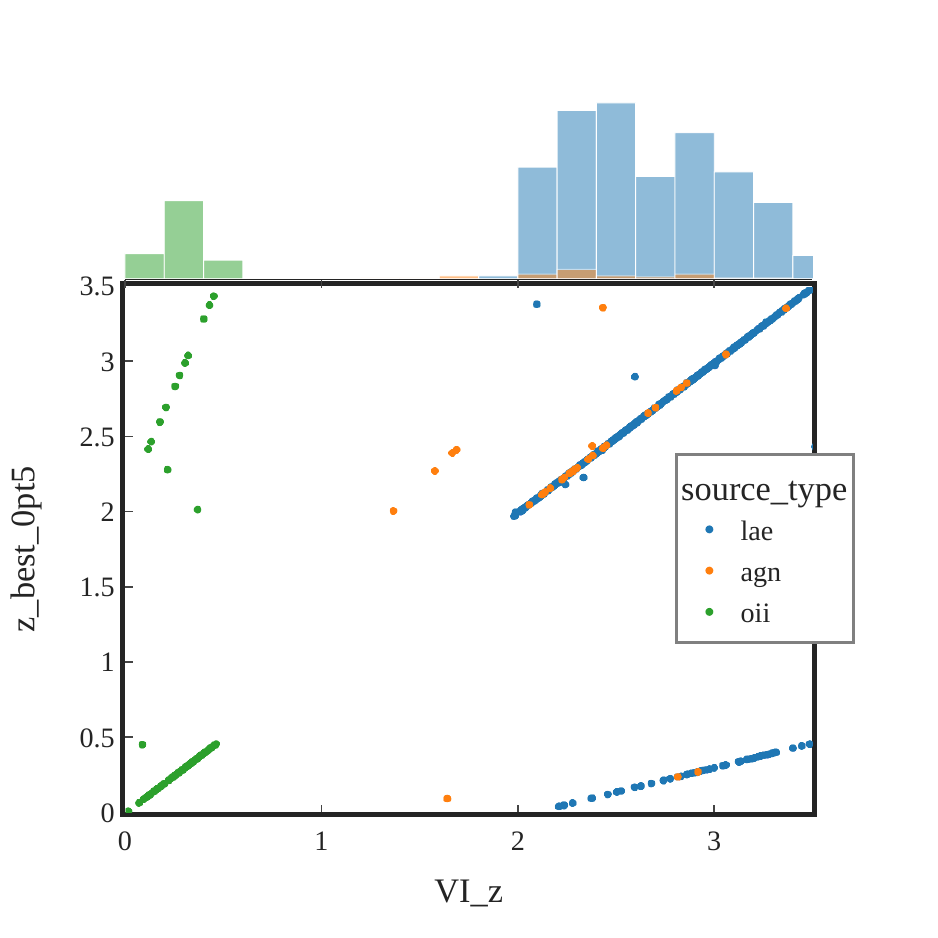}
  \caption{Comparison of redshifts obtained from visual inspection, \texttt{VI\_Z}, of DESI spectra ($x$-axis) and those derived from HETDEX data ($y$-axis) using ELiXer. The plot assumes a P(\lya)=0.5 threshold for the HETDEX line classifications. As discussed in Section~\ref{sec:classification}, this value leads to a 1.15\%  contamination of the LAE sample by [\ion{O}{2}] galaxies, and a loss of 5.6\% true LAEs from the HETDEX sample.  }\label{fig:zcomp}
\end{figure}

Figure~\ref{fig:zcomp} shows a comparison between redshifts based on the HETDEX classification code using the P(\lya)=0.5 threshold and redshifts  determined from the visual inspection of DESI spectra.  This value of  P(\lya) represents a trade-off between \OII contamination and \lya recovery: lowering its value increases the rate of LAE recovery (i.e., it reduces the number of points in the bottom-right region of the figure) at the expense of greater contamination (adding points in the upper left). Of the single broad-line AGN classified by ELiXer as \lya, 8 were shown by DESI spectra to be \ion{C}{4} $\lambda 1549$ based on the presence of secondary lines such as \ion{C}{3}] $\lambda 1909$ and \ion{Mg}{2} $\lambda 2800$.  Both these lines fell outside the HETDEX wavelength range.

\begin{table}
\begin{center} \begin{tabular}{c|cc}
     P(\lya) Threshold  & [\ion{O}{2}] Contamination   & \lya Recovery \\
     \hline
 0.9 & 0.97\% & 91.4\% \\
 0.8 & 0.97\% & 91.5\% \\
 0.7 & 0.97\% & 91.6\% \\
 0.6 & 0.97\% & 91.7\% \\
 \textbf{0.5} & \textbf{1.15\%} & \textbf{94.4\%} \\
 0.4 & 2.13\% & 95.8\% \\
 0.3 & 2.71\% & 96.2\% \\
 0.2 & 3.72\% & 97.6\% \\
 0.1 & 6.67\% & 98.6\% \\

     \end{tabular}
   \end{center}
   \caption{Summary of the results of HETDEX ELiXer classification of visually vetted spectra, see Figure \ref{fig:elixer_plya}. The Contamination and Recovery columns for each row are independently computed using that row's P(\lya) Threshold. The row in \textbf{boldface} shows the default ELiXer classification threshold configuration. The results are consistent with those in \cite{davis2023} despite differences in the distributions of the magnitudes and emission line identifications. }
   \label{tab:elixer_plya}
   \end{table}

\begin{table*}
\centering
\caption{Classification and line fit parameters. The first 10 rows of the 2374 row catalog associated with this paper are printed for demonstrative purposes. The full version can be found online. In the Appendix, a full description of the FITS file released with this paper can be found found in Table~\ref{tab:hdu_info} with a description of each column described in Table~\ref{tab:columns}. All wavelength values are reported in air, in units of \AA\null. Line width values, $\sigma$, are reported in \AA\null. The HETDEX values are not corrected for instrumental broadening. The flux values reported are in \fluxunit . \label{tab:cat}}
\begin{tabular}{cccccccc}
\hline \hline
TARGETID & TARGET\_RA & TARGET\_DEC & TILEID & RA\_HETDEX & DEC\_HETDEX & SEP & DETECTID \\
\hline
103606502031361 & 224.490280 & 50.115120 & 80870 & 224.4905 & 50.1151 & 0.4960 & 3007922753 \\
103606506225670 & 224.752151 & 50.046459 & 80870 & 224.7521 & 50.0464 & 0.3578 & 3007918479 \\
103606506225674 & 224.588257 & 50.011024 & 80870 & 224.5883 & 50.0110 & 0.1413 & 3007918678 \\
103606506225675 & 224.637772 & 50.085388 & 80870 & 224.6378 & 50.0854 & 0.2018 & 3007919753 \\
103606506225676 & 224.721085 & 50.040512 & 80870 & 224.7212 & 50.0406 & 0.2334 & 3007918756 \\
103606506225680 & 224.657257 & 50.085510 & 80870 & 224.6575 & 50.0855 & 0.4738 & 3007919755 \\
103606506225682 & 224.762589 & 50.021271 & 80870 & 224.7626 & 50.0212 & 0.2984 & 3007918514 \\
103606506225685 & 224.833771 & 50.057350 & 80870 & 224.8338 & 50.0574 & 0.0017 & 3007918908 \\
103606506225686 & 224.667603 & 50.055523 & 80870 & 224.6678 & 50.0555 & 0.4585 & 3007919332 \\
103606506225687 & 224.779495 & 50.102844 & 80870 & 224.7795 & 50.1028 & 0.0045 & 3007921868 \\
\hline
\end{tabular}

\vspace{4pt}
\resizebox{\textwidth}{!}{
\begin{tabular}{cccccccccc}
\hline \hline
VI\_Z & VI\_QUALITY & SOURCE\_TYPE & DEX\_FLAG & Z\_BEST\_0PT3 & Z\_BEST\_0PT4 & Z\_BEST\_0PT5 & AC & GMAG & SN\_HETDEX \\
\hline
-1.0000 & 0 & lae & 1 & 2.2106 & 2.2106 & 2.2106 & 0.0469 & 25.8217 & 5.9500 \\
3.4182 & 3 & lae & 1 & 3.4182 & 3.4182 & 3.4182 & 0.0518 & 24.8484 & 5.3700 \\
-1.0000 & 0 & lae & 1 & 2.3794 & 2.3794 & 2.3794 & 0.0526 & 25.0488 & 5.1200 \\
-1.0000 & 0 & lae & 0 & 2.6206 & 2.6206 & 2.6206 & 0.0530 & 24.0520 & 5.3300 \\
2.9116 & 3 & lae & 1 & 0.2756 & 0.2756 & 0.2756 & 0.0528 & 22.9939 & 6.8700 \\
-1.0000 & 0 & lae & 1 & 2.2066 & 2.2066 & 2.2066 & 0.0536 & 25.4402 & 4.8692 \\
3.0000 & 3 & lae & 1 & 2.9997 & 2.9997 & 2.9997 & 0.0514 & 26.0349 & 7.6200 \\
-1.0000 & 0 & oii & 1 & 0.4678 & 0.4678 & 0.4678 & 0.0499 & 23.6870 & 5.2800 \\
-1.0000 & 0 & lae & 1 & 2.1348 & 2.1348 & 2.1348 & 0.0539 & 24.8588 & 5.8300 \\
2.7621 & 4 & lae & 1 & 2.7623 & 2.7623 & 2.7623 & 0.0496 & 26.4235 & 6.2300 \\
\hline
\end{tabular}}

\vspace{4pt}
\resizebox{\textwidth}{!}{
\begin{tabular}{cccccccc}
\hline \hline
WAVE\_HETDEX & WAVE\_ERR\_HETDEX & FLUX\_HETDEX & FLUX\_ERR\_HETDEX & SIGMA\_HETDEX & SIGMA\_ERR\_HETDEX & CONT\_HETDEX & 
CONT\_ERR\_HETDEX \\
\hline
3901.9700 & 0.6000 & 13.9400 & 3.1300 & 1.8600 & 0.5800 & -0.1400 & 0.1800 \\
5369.7100 & 0.5100 & 7.8600 & 1.4600 & 1.9700 & 0.5100 & 0.0050 & 0.1000 \\
4107.1602 & 0.8300 & 14.5900 & 2.9500 & 3.5400 & 0.7600 & 0.0450 & 0.1500 \\
4400.2500 & 0.4900 & 8.3600 & 2.1800 & 1.6500 & 0.6500 & 0.0650 & 0.1300 \\
4755.1802 & 0.6500 & 15.4400 & 2.1400 & 4.4000 & 0.7400 & 0.1850 & 0.1000 \\
3897.0801 & 0.5600 & 12.1500 & 2.8200 & 2.0200 & 0.7000 & -0.1300 & 0.1800 \\
4861.0498 & 0.3900 & 16.0500 & 1.4600 & 3.8000 & 0.4000 & 0.0050 & 0.0800 \\
5471.7100 & 0.4000 & 8.1600 & 1.2100 & 2.0800 & 0.4000 & 0.0650 & 0.0800 \\
3809.8501 & 0.2900 & 15.3600 & 1.9700 & 1.8100 & 0.2700 & -0.1500 & 0.1500 \\
4572.5400 & 0.5600 & 11.6800 & 1.7800 & 2.9600 & 0.5200 & -0.0300 & 0.1000 \\
\hline
\end{tabular}}

\vspace{4pt}
\resizebox{\textwidth}{!}{
\begin{tabular}{cccccccc}
\hline \hline
WAVE\_DESI & WAVE\_ERR\_DESI & FLUX\_DESI & 
FLUX\_ERR\_DESI & SIGMA\_DESI & SIGMA\_ERR\_DESI & CONT\_DESI & CONT\_ERR\_DESI \\
\hline
3901.0369 & 3.7251 & 0.9410 & 0.8899 & 6.0968 & 3.4219 & -0.0795 & 0.0239 \\
5369.4976 & 0.1637 & 8.3260 & 0.5096 & 1.9146 & 0.1712 & 0.0192 & 0.0108 \\
4104.7891 & 3.3367 & 0.8758 & 0.7595 & 5.7900 & 3.1882 & 0.0095 & 0.0174 \\
4398.3242 & 3.5279 & 0.7357 & 0.6811 & 5.8236 & 3.2706 & -0.0131 & 0.0155 \\
4754.4004 & 0.2762 & 10.2795 & 0.7797 & 2.7820 & 0.3651 & 0.1109 & 0.0127 \\
3895.1643 & 3.6269 & 0.6176 & 0.6168 & 5.4959 & 3.5032 & 0.0786 & 0.0236 \\
4861.7876 & 0.3043 & 12.7067 & 0.7819 & 3.7228 & 0.3496 & -0.0088 & 0.0113 \\
5470.2783 & 3.6197 & 0.6026 & 0.5033 & 5.9930 & 3.1566 & -0.0087 & 0.0111 \\
3806.3154 & 3.5183 & 1.4814 & 1.4153 & 6.2903 & 3.2750 & -0.1125 & 0.0326 \\
4572.7666 & 0.4182 & 7.2106 & 0.9677 & 2.6546 & 0.6306 & 0.0425 & 0.0152 \\
\hline
\end{tabular}}

\end{table*}

\section{Emission line comparisons}
\label{sec:linecomps}

In this section, we compare HETDEX and DESI measurements of emission-line flux and line width of the confirmed LAEs and \OII\ emitters with \texttt{VI\_QUALITY} $\ge$ 3. Any emission lines blueward of 3800~\AA\ are excluded due to lower $S/N$ in the DESI spectra.

To minimize systematic errors in our line measurements, we use the same algorithms for both groups of spectra.  The input to the line-fitting algorithm are the DESI and HETDEX spectra, reduced by their respective collaboration pipelines; for HETDEX the pipeline used was that associated with the internal HDR3.  After the reduction, the analyses for both spectra were identical. 

We model the line emission using the software package \texttt{mcmc\_gauss.py}, developed within HETDEX's software repository, \texttt{hetdex-api}\footnote{\url{https://github.com/HETDEX/hetdex_api/blob/master/hetdex_tools/mcmc_gauss.py}}. In this method, a single Gaussian model is fit to the spectral line using the publicly available \texttt{emcee}\footnote{\texttt{\url{https://emcee.readthedocs.io/}}}\citep{emcee}, which implements Goodman \& Weare’s Affine Invariant Markov chain Monte Carlo (MCMC) Ensemble sampler.  For the initial guess, we use the wavelength, flux, and line width as determined by the HETDEX pipeline.  For determining the line properties of the DESI spectra, we limit the range to $\pm 50$~\AA\ of the known line position. This is sufficient to obtain reliable fits for the majority of targets.

\begin{figure}[t]
    \centering    \includegraphics[width=0.48\textwidth]{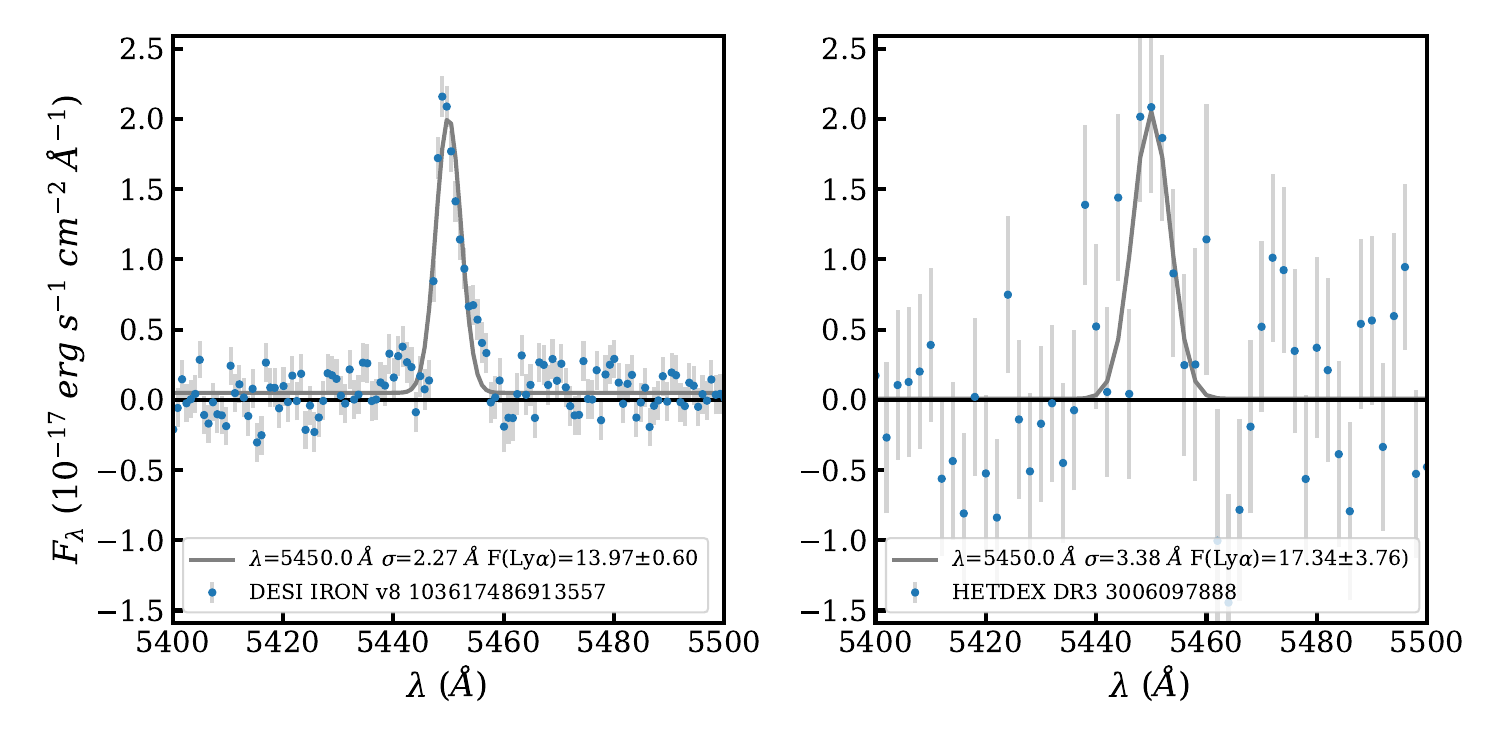}
    \includegraphics[width=0.48\textwidth]{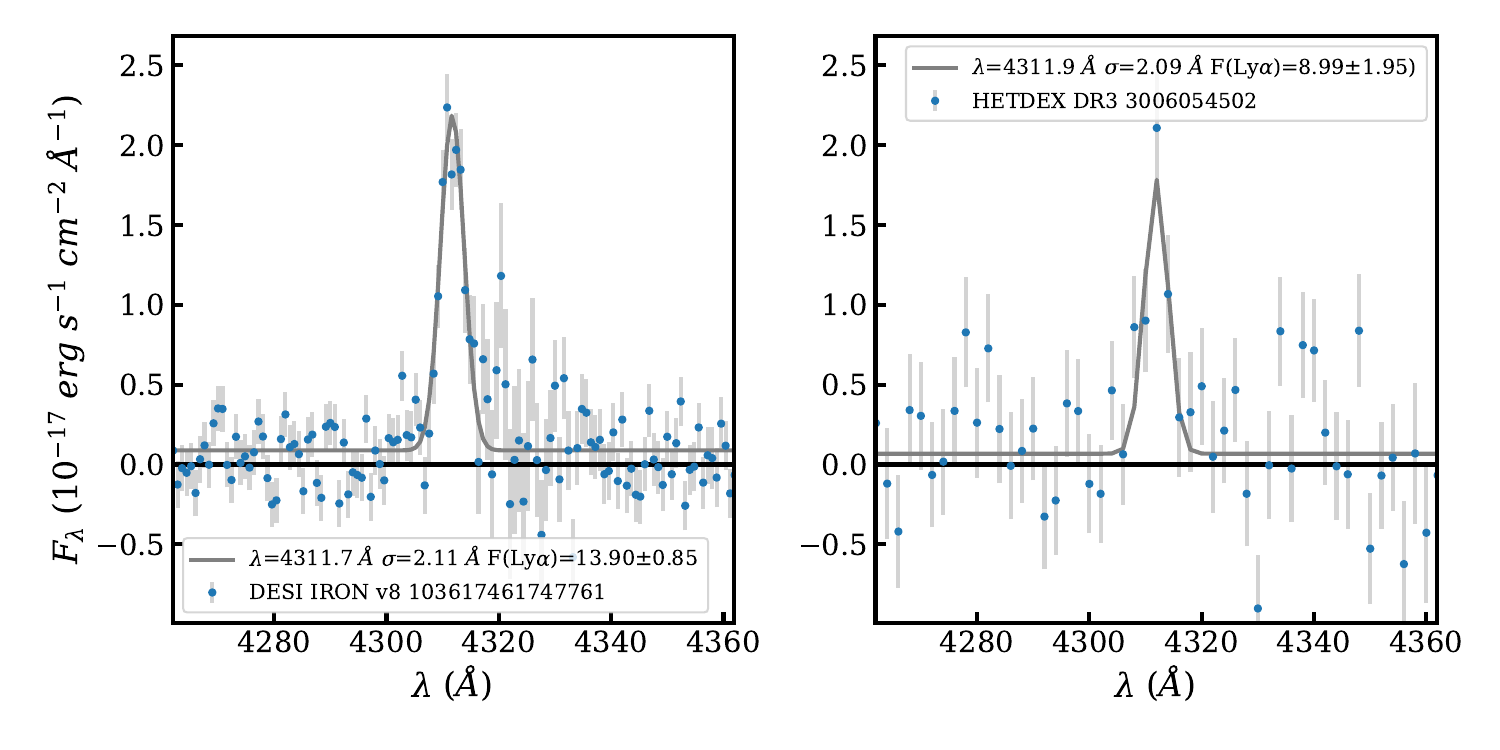}
    \includegraphics[width=0.48\textwidth]{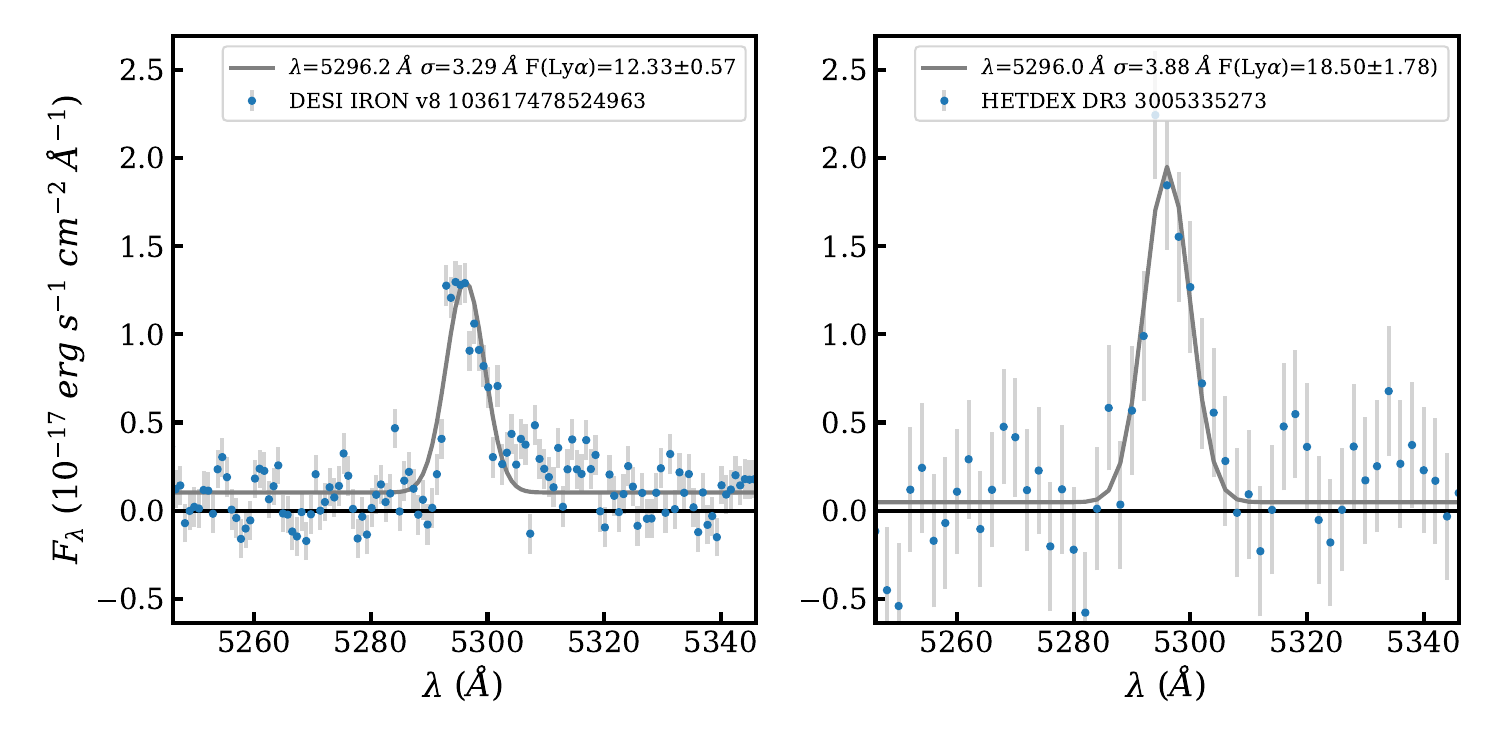}
        \includegraphics[width=0.48\textwidth]{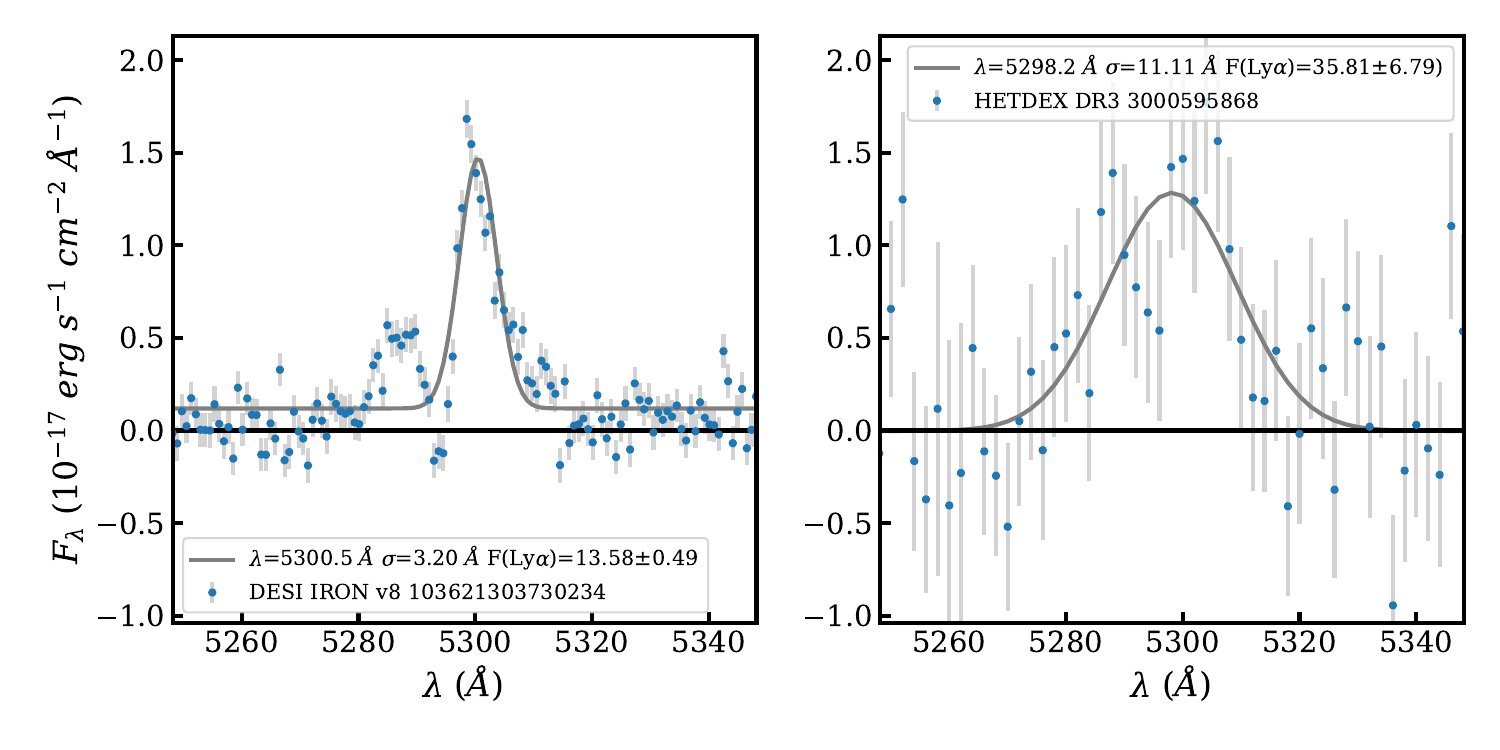}
    \caption{Examples of fitted \lya lines for DESI on left and HETDEX on right.  Each row represents a single LAE\null.  The \lya flux values in the legends are reported in units of \fluxunit\null. In the bottom example, the line is double peaked; as a result, the fitted linewidth is higher for HETDEX than for the DESI line fit which only fits the brightest component of the line.}
    \label{fig:line_fit_ex}
\end{figure}

Figure~\ref{fig:line_fit_ex} shows an example of four fitted lines of LAEs. The left column shows the higher spectral resolution and more sensitive DESI spectroscopy; the right column displays the corresponding spectra obtained by HETDEX\null.
The $y$-axis is fixed to the same scale limits to demonstrate differences in absolute flux. For extended sources, fluxes can vary significantly, as the HETDEX spectra are derived from an IFU while DESI only captures light from a single $1\farcs 5$ fiber, and no aperture corrections for extended emission are applied. 

The bottom example shows a double-peaked \lya line profile. With the higher spectral resolution of DESI, a single Gaussian model is a poor fit to the total emission. In contrast, the HETDEX spectrum with its higher noise, fits both peaks to a single Gaussian and consequently obtains a higher line width, $\sigma_{\mathrm{HETDEX}}$.

\begin{figure}[t]
\centering

\includegraphics[trim={0.2cm 0.1cm 0.1cm 0.2cm},clip,width=0.45\textwidth]{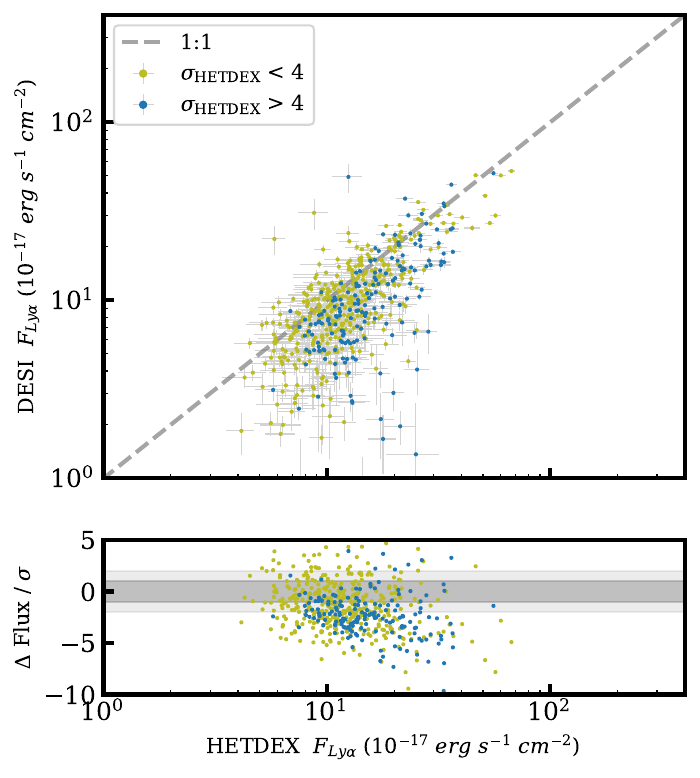}
\caption{Line flux comparison for the 982 LAEs detected by HETDEX confirmed by DESI (\texttt{VI\_QUALITY}$ \ge 3$) and the error--weighted difference is shown in the bottom panel. Data points are separated based on HETDEX line width values, \texttt{SIGMA\_HETDEX}. As demonstrated in the bottom panel of Figure~\ref{fig:line_fit_ex}, asymmetric and double-peaked line profiles are fit with higher line widths in the HETDEX spectra resulting in higher Ly$\alpha$ flux values.}
    \label{fig:linefluxcomp}
\end{figure}

\subsection{Line Fluxes}
Figure~\ref{fig:linefluxcomp} shows the ensemble comparison of \lya line fluxes between the two surveys. In the bottom panel, we show error-weighted difference between the two measures as a function of HETDEX flux values ($(F_{DESI}-F_{HETDEX})/\sqrt{\sigma^2_{DESI}+\sigma^2_{HETDEX}}$). Generally, HETDEX flux values are found to be higher than the DESI values, with the difference growing larger with increased flux. Since no aperture corrections are applied to the DESI data, this offset is not surprising. At higher line fluxes, many LAEs are extended and flux will be lost when observed with a single fiber. In addition, in some cases where the line is double-peaked, the higher spectral resolution DESI data only fits one peak and a significant fraction of the \lya line flux is lost.  This latter effect is illustrated in the bottom panel of Figure~\ref{fig:line_fit_ex}.

A significant contribution to the scatter also comes from centering uncertainty. As discussed in Section~\ref{sec:missing}, simulations show that the ability for the HETDEX detection pipeline to find the true position of a source decreases in accuracy as $S/N$ decreases. For robust emission line sources detected at $S/N>5.5$, up to 10\% of sources are found more than 1\arcsec\ away from the input position. Lower $S/N$ emission lines are much more difficult to locate their precise center. Over 20\% of simulated sources have output positions over 1\arcsec~away from their input position. With just a single DESI fiber placed on the source, the position uncertainty translates to a flux loss in the target DESI fiber.

Positional uncertainty contributes to the scatter in the measured line fluxes between the two surveys and it also contributes to large scatter in the measured spectral continuum. Using spectral data from both surveys, a common pseudo-magnitude value can be derived for both datasets by convolving a $g$-band filter curve with the spectral data. On average the HETDEX magnitudes are brighter ($\langle$ \hetg\ - $g_{\mathrm{DESI}} \rangle$ = -0.66~mag ) and exhibit an rms scatter of $1.15$\,mag. Generally, the continuum flux of HETDEX LAEs are too faint to be detected by VIRUS which reaches a sensitivity of \hetg$\sim25$~mag so in most cases the HETDEX continuum measures are an upper limit.

\subsection{Wavelengths and Redshifts}
\begin{figure}
    \centering
    \includegraphics[trim={0.2cm 0.1cm 0.1cm 0.2cm},clip,width=0.95\linewidth]{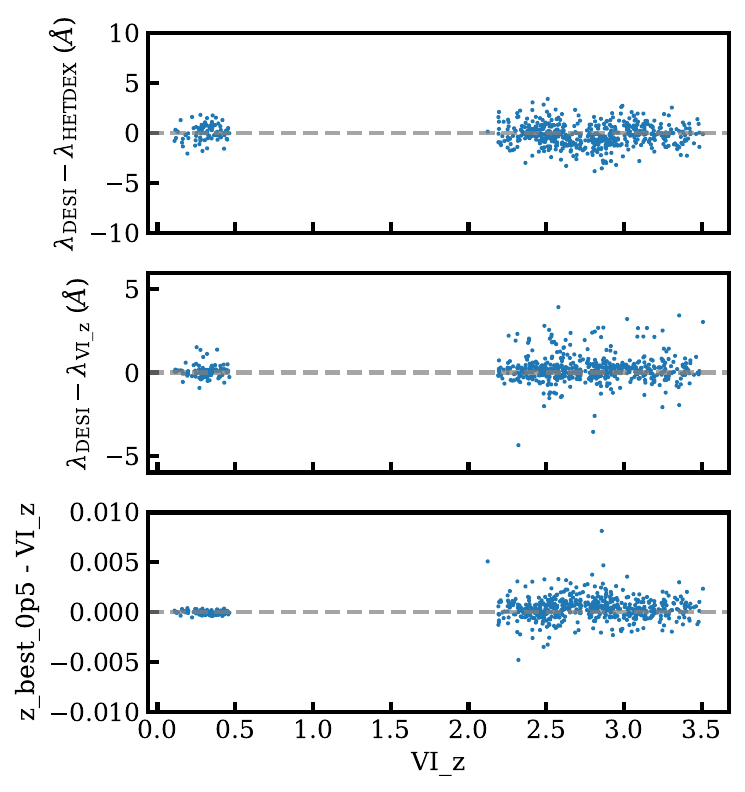}
    \caption{A comparison between the best-fit central wavelengths calculated for each experiment. Both use a single Gaussian model and are fit with their wavelengths as measured in air. The middle panel converts the visual spectroscopic redshift, $\texttt{VI\_Z}$, to a Ly$\alpha$, or \OII wavelength value redshifted to the observed frame and compares to the best fit wavelength in the DESI data. The bottom panel presents a comparison between HETDEX spectroscopic redshifts and DESI VI spectroscopic redshifts.}
    \label{fig:waves}
\end{figure}

A comparison between the best-fit central wavelengths calculated for each experiment is shown in Figure~\ref{fig:waves}. DESI spectral data are converted to wavelength measurements in air to match those of HETDEX. Values are derived from line fitting each spectrum using a single Gaussian model. The two values agree very well between the two surveys with an average offset of $\langle \lambda \rangle = 0.15$\,\AA \ and a rms scatter of 1.17\AA. 

We can also consider a comparison to the best-fit visual wavelength. During the VI process, the classifier tunes the location of the Ly$\alpha$ line. The classifier must decide whether they are fitting to the peak of the Ly$\alpha$ emission or instead to the middle of the emission when an asymmetric profile shape is clearly seen. This decision will lead in some scatter relative to the wavelength determined from automated fitting and consequently some scatter in the spectroscopic redshift determined for the LAE. The middle panel in Figure~\ref{fig:waves} converts the visual spectroscopic redshift, $\texttt{VI\_Z}$, to a Ly$\alpha$ wavelength value in the observe frame and compares to the best-fit DESI wavelength. These values are consistent within the scatter of 1.37~\AA\null. The dispersion in the determined wavelength is higher in this case relative to the above comparison between the automated values from the two surveys.

The bottom panel presents a comparison between HETDEX spectroscopic redshifts and DESI VI spectroscopic redshifts. The HETDEX redshifts are the result of the classifcation pipeline described in Section~\ref{sec:classification} and come from fitting a single Gaussian model to the emission line.
The DESI redshifts are from visually tuning to the peak emission wavelength using the DESI VI tool. Ignoring gross misclassifications, \textit{i.e.} those sources that lie $\Delta z > 0.1$ from the main diagonal in figure~\ref{fig:zcomp}, the median difference between the two values is $\langle \Delta z / (1 + z) \rangle = 6.87 \times 10^{-5}$ with an rms scatter of $3.32 \times 10^{-4}$.

\subsection{Linewidths}

\begin{figure}
\centering
\includegraphics[trim={0.4cm 0.1cm 0.1cm 0.2cm},clip,width=0.45\textwidth]{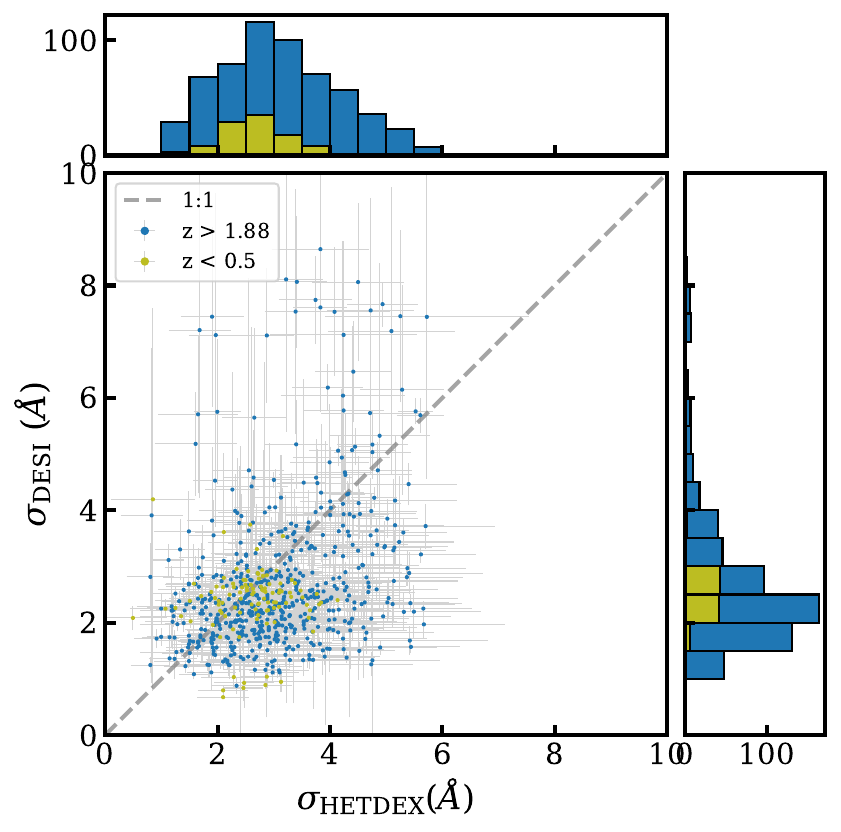}

    \caption{Comparison of Gaussian line width, $\sigma$, values between DESI ($y$-axis) and HETDEX ($x$-axis). Both line profiles are modeled using a single Gaussian model. HETDEX values are corrected for the lower instrumental resolution of the VIRUS spectrograph compared to the DESI spectrograph. The values show large scatter at higher line width where the difference in spectral resolution can lead to fitting of only one peak in the double-peaked and asymmetric lines found in the LAE sample. }
    \label{fig:linewidthcomp}
\end{figure}

Figure~\ref{fig:linewidthcomp} presents a comparison between the Gaussian line widths measured by each survey. The HETDEX value, $\sigma_\mathrm{HETDEX}$, is corrected for the VIRUS instrument's low spectral resolution ($ R\sim800$). The values reported in the data table of this paper contain the measured, uncorrected values. Once corrected, the line width values agree. 

While there is some scatter, the two projects agree well about the basic properties of the detected emission lines.  The flux and line width ratios are within 1-$\sigma$ of 1 and the error weighted differences are within 1 standard deviation of 0.

\begin{equation}
\begin{array}{rcl}
\frac{flux (DESI))}{flux (HETDEX)} &=& 0.853 \pm 0.391\\
& & \\
\frac{flux (DESI) - flux (HETDEX)}{\sqrt{\sigma_{f,~ DESI}^2 + \sigma_{f, ~ HETDEX}^2}} &=&
-1.023 \pm 2.441\\
& & \\
\frac{width (DESI)}{width (HETDEX)} &=& 0.859 \pm 0.550\\
& & \\
\frac{width (DESI) - width (HETDEX)}{\sqrt{\sigma_{w,~ DESI}^2 + \sigma_{w,~ HETDEX}^2}} &=&
-0.71 \pm 1.79
\end{array}
\end{equation}

\section{Conclusion}\label{sec:conclusion}

Observations of HETDEX LAE candidates with DESI has enabled the validation of the
HETDEX line classification method in separating \lya from
[\ion{O}{2}] emitters and demonstrates that the HETDEX survey achieves the science requirement of less than 2\% interloper contamination. While the DESI sample is biased to both higher continuum fluxes and higher S/N line fluxes (these correlate), in both the case of lower S/N (where we have poorer centering accuracy) and lower continuum flux, the confusion between LyA and OII is minimal -- they are all classified as LAEs.  So compared to the internal HETDEX LAE sample, the DESI sample would have a higher contamination rate. For those sources that are confidently detected in DESI, the derived central wavelength of the emission line, their line fluxes and line widths are in good agreement between the two experiments. The spectroscopic redshifts obtained from each survey, when the classifications agree, are in very good agreement with only a small median offset between the two surveys of $\langle \Delta z / (1 + z) \rangle = 6.87 \times 10^{-5}$ with an rms scatter of $3.32 \times 10^{-4}$.

\section*{Acknowledgements}

This material is based upon work supported by the U.S. Department of Energy (DOE), Office of Science, Office of High-Energy Physics, under Contract No. DE–AC02–05CH11231, and by the National Energy Research Scientific Computing Center, a DOE Office of Science User Facility under the same contract. Additional support for DESI was provided by the U.S. National Science Foundation (NSF), Division of Astronomical Sciences under Contract No. AST-0950945 to the NSF’s National Optical-Infrared Astronomy Re- search Laboratory; the Science and Technologies Facilities Council of the United Kingdom; the Gordon and Betty Moore Foundation; the Heising-Simons Foundation; the French Alternative Energies and Atomic Energy Commission (CEA); the National Council of Sci- ence and Technology of Mexico (CONACYT); the Ministry of Science and Innovation of Spain (MICINN), and by the DESI Member Institutions: https://www.desi. lbl.gov/collaborating-institutions. Any opinions, findings, and conclusions or recommendations expressed in this material are those of the author(s) and do not necessarily reflect the views of the U.S. National Science Foundation, the U.S. Department of Energy, or any of the listed funding agencies.

The authors are honored to be permitted to conduct scientific research on Iolkam Du’ag (Kitt Peak), a mountain with particular significance to the Tohono O’odham Nation.

HETDEX is led by the University of Texas at the Austin McDonald
Observatory and Department of Astronomy with participation from the
Ludwig-Maximilians-Universit\"{a}t M\"{u}nchen, Max-Planck-Institut für
Extraterrestrische Physik (MPE), Leibniz-Institut f\"{u}r Astrophysik
Potsdam (AIP), Texas A\&M University, The Pennsylvania State
University, Institut für Astrophysik G\"{o}ttingen, The University of
Oxford, Max- Planck-Institut f\"{u}r Astrophysik (MPA), The University of
Tokyo, and Missouri University of Science and Technology. In addition to institutional support, HETDEX is funded by the
National Science Foundation (grant AST-0926815), the State of Texas,
and the US Air Force (AFRL FA9451-04-2-0355) and receives generous
support from private individuals and foundations. 

The observations were obtained with the Hobby-Eberly Telescope (HET), which is a joint project of the University of Texas at Austin, the Pennsylvania State University, Ludwig-Maximilians-Universität München, and Georg-August-Universität Göttingen. The HET is named in honor of its principal benefactors, William P. Hobby and Robert E. Eberly.

The authors acknowledge the Texas Advanced Computing Center (TACC) at The University of Texas at Austin for providing high performance computing, visualization, and storage resources that have contributed to the research results reported within this paper. URL: http://www.tacc.utexas.edu

\bibliographystyle{aasjournal}
\bibliography{refs}
\facility{KPNO:Mayall (DESI)}
\facility{McDonald:HET (VIRUS)}

\newpage
\appendix

\label{appendix:1}

\begin{table}[h]
\caption{FITS File HDU Information for \texttt{DESI\_HETDEX\_SPEC\_v1.6.fits}}
\label{tab:hdu_info}
\centering
\begin{tabular}{l|l|l|l|l}
\hline
\hline
No. & Name & Type & Dimensions & Column Description \\
\hline
0  & PRIMARY           & PrimaryHDU  & ()              &  \\
1  & INFO              & BinTableHDU & 2374R $\times$ 34C & Main Table. Columns described in Table~\ref{tab:columns} \\
2  & HETDEX\_WAVE      & ImageHDU    & (1036)          & Wavelength array, in air, for HETDEX data in \AA\ \\
3  & HETDEX\_SPEC      & ImageHDU    & (1036, 2374)    & Spectral flux array for HETDEX data in $1 \times 10^{-17} \; \mathrm{erg\,\mathring{A}^{-1}\,s^{-1}\,cm^{-2}}$ \\
4  & HETDEX\_SPEC\_ERR & ImageHDU    & (1036, 2374)    & Uncertainty in HETDEX spectral flux in $1 \times 10^{-17} \; \mathrm{erg\,\mathring{A}^{-1}\,s^{-1}\,cm^{-2}}$ \\
5  & DESI\_WAVE        & ImageHDU    & (7781)          & Wavelength array, in air, for DESI data in \AA\ \\
6  & DESI\_WAVE\_VACUUM & ImageHDU   & (7781)          & Wavelength array (in vacuum) for DESI data in \AA\ \\
7  & DESI\_SPEC        & ImageHDU    & (7781, 2374)    & Spectral flux array for DESI data in $1 \times 10^{-17} \; \mathrm{erg\,\mathring{A}^{-1}\,s^{-1}\,cm^{-2}}$ \\
8  & DESI\_SPEC\_ERR   & ImageHDU    & (7781, 2374)    & Uncertainty in DESI spectral flux in $1 \times 10^{-17} \; \mathrm{erg\,\mathring{A}^{-1}\,s^{-1}\,cm^{-2}}$ \\
\hline
\end{tabular}
\end{table}

\begin{table}[h]
\caption{Column Description for INFO BinTable in the HETDEX-DESI Catalog}
    \centering
    \begin{tabular}{l|l|l}
\hline
\hline
Column Name &  Column Description & Data Type \\
\hline
TARGETID     & DESI ID       & int64  \\ 
TARGET\_RA     & DESI fiber position right ascension (ICRS deg)       & float64  \\ 
TARGET\_DEC     & DESI fiber position declination (ICRS deg)       & float64  \\ 
TILEID     & DESI Tile ID       & int32  \\ 
RA\_HETDEX     & HETDEX detectid right ascension in HDR3 (ICRS deg)       & float32  \\ 
DEC\_HETDEX     & HETDEX detectid declination in HDR3 (ICRS deg)       & float32  \\ 
SEP     & separation between HETDEX detectid and the DESI fiber position       & float32  \\ 
DETECTID     & HETDEX emission line detection ID       & int64  \\ 
VI\_Z     & Visual spectroscopic redshift       & float32  \\ 
VI\_QUALITY     & Quality in VI\_z (4=multiple spectral features, 3=single emission line)       & int32  \\ 
SOURCE\_TYPE     & HETDEX source\_type (lae, oii, agn, other)       & str10  \\ 
DEX\_FLAG     & HETDEX detections that are now flagged as compromised, ignored in analysis       & int64  \\ 
Z\_BEST\_0PT3     & redshift determined from ELiXeR with threshold (P(\lya)=0.3)       & float32  \\ 
Z\_BEST\_PT4     & redshift determined from ELiXeR with threshold (P(\lya)=0.4)       & float32  \\ 
Z\_BEST\_0PT5     & redshift determined from ELiXeR with threshold (P(\lya)=0.5)       & float32  \\ 
AV     & dust correction in V band -- not applied to fluxes       & float32  \\ 
GMAG     & sdss-g magnitude measured in HETDEX spectrum       & float32  \\ 
SN\_HETDEX     & HETDEX signal-to-noise for line emission       & float32  \\ 
WAVE\_HETDEX     & central wavelength of line emission (\AA, in air)       & float32  \\ 
WAVE\_ERR\_HETDEX     & mcmc error in wave       & float32  \\ 
FLUX\_HETDEX     & observed line flux at WAVE\_HETDEX ($1 \times 10^{-17} \; \mathrm{erg\,s^{-1}\,cm^{-2}}$)       & float32  \\ 
FLUX\_ERR\_HETDEX     & mcmc error in line FLUX\_HETDEX ($1 \times 10^{-17} \; \mathrm{erg\,s^{-1}\,cm^{-2}}$)       & float32  \\ 
SIGMA\_HETDEX     & sigma linewidth in gaussian line fit (\AA)       & float32  \\ 
SIGMA\_ERR\_HETDEX     & mcmc error in sigma linewidth (\AA)       & float32  \\ 
CONT\_HETDEX     & local fitted observed continuum       & float32  \\ 
CONT\_ERR\_HETDEX     & mcmc error in continuum       & float32  \\ 
WAVE\_DESI     & central wavelength of line emission (\AA, in air)       & float32  \\ 
WAVE\_ERR\_DESI     & mcmc error in wave       & float32  \\ 
FLUX\_DESI     & observed line flux at WAVE\_DESI ($1 \times 10^{-17} \; \mathrm{erg\,s^{-1}\,cm^{-2}}$)       & float32  \\ 
FLUX\_ERR\_DESI     & mcmc error in line FLUX\_DESI ($1 \times 10^{-17} \; \mathrm{erg\,s^{-1}\,cm^{-2}}$)       & float32  \\ 
SIGMA\_DESI     & sigma linewidth in gaussian line fit (\AA)       & float32  \\ 
SIGMA\_ERR\_DESI     & mcmc error in sigma linewidth (\AA)       & float32  \\ 
CONT\_DESI     & local fitted observed continuum       & float32  \\ 
CONT\_ERR\_DESI     & mcmc error in continuum       & float32  \\ 
\hline
\end{tabular}
    \label{tab:columns}

\end{table}

\end{document}